%% file: moriond09.tex
\documentclass[11pt]{article}
\usepackage{moriond,epsfig,amssymb}
\usepackage{macros_static}
\usepackage{wrapfig}

\def\platz{\hspace*{5mm}}

\begin{document}
\vspace*{4cm}
\begin{flushright}
\vskip -2.7cm
DESY 09-088 \\
SFB/CPP-09-48\\[2ex]
\end{flushright}
\title{\MakeUppercase{
New perspectives for heavy flavour physics from the
lattice}}

\author{ R. SOMMER }

\address{NIC, DESY, Platanenallee 6, 15738 Zeuthen, Germany}

\maketitle\abstracts{
Heavy flavours represent a challenge for lattice QCD. We discuss it
in very general terms. 
We give an idea of the significant recent progress which opens up 
good perspectives for high precision first principles 
QCD computations for flavour physics.
}

\section{Overview}\label{s:overview}
In this talk I want to convey an idea about 
the perspectives for precise lattice QCD computations. 
Some emphasis is put on the field of flavour physics, where
lattice QCD seems to be needed the most in the quantitative
interpretation of present and future experiments (in particular this
is the reason for \sect{s:beauty}). 
I do not want to discuss the latest numbers but will 
focus on the principle, the challenges and the perspectives.
I mainly address those in the audience who know little about the field
and therefore take a bit of a bird's-eye view of the field.
So I do not hesitate to also list trivial facts 
(e.g. \sect{s:machines})
and my personal opinion on what is relevant for the future. 

\section{The principle} \label{s:principle}

We want to find answers for properties of QCD in the
non-perturbative domain, where the usual power series
expansion in the coupling constant fails completely. This is
done through a \\[0.5ex]
\platz  Monte Carlo (MC) evaluation\\[0.5ex]
\platz  of the Euclidean path integral\\[0.5ex]
\platz after a discretization of space-time 
  on a lattice with spacing $a$ in all 3+1 
  dimensions.\\[1ex]
The last of these items seems like a bad mutilation of the
theory, but we emphasize that
discretization is just one form of regularization,
i.e. the taming of the infamous UV divergences of 
continuum path integrals. In fact discretization is the
only known, complete regularization of the theory: complete in
the sense that it defines the theory mathematically
also beyond an expansion in the coupling.

In lattice QCD the {\em Euclidean} path integral 
is ``solved'' numerically. Many observables can
be computed directly in Euclidean space-time, but 
even more cannot be computed (unless the {\em complete} 
dependence on time is known). Examples are scattering cross sections.

What at first sight looks like the (classical)
coupling constant, $g_0$, and quark masses, $m_{0,i}$, have no direct
physical meaning in a quantum field theory such as QCD. 
They are bare (actually divergent) quantities which need
renormalization. In the general (non-perturbative) case, renormalization 
just means that the bare parameters
are replaced by observable quantities. It is convenient to 
parametrize the theory as follows. The MC evaluation of the
path integral yields dimensionless hadron masses (remember that $a$
is the lattice spacing), 
\bes
  a m^\mrm{had}_h = \mu^\mrm{had}_h(g_0,\{\,\mu^\mrm{quark}_i\,\})
  \label{e:masses}
\ees
as a function of the bare, dimensionless,  parameters in the
Lagrangian of the theory,
$g_0,\;\mu^\mrm{quark}_i=am^\mrm{quark}_{0,i}\,,\;i=\up,\down,\ldots$. 
We consider the theory 
with $\nf$ quarks and require $\nf$ ratios of hadron masses
to agree with Nature. This may be read as fixing the
bare dimensionless quark masses $\mu_i$ to the correct values
at a given $g_0$.
Taking now in addition the dimensionfull masse(s) $m_h$
from experiment, yields the lattice spacing
\bes
  a = {\mu^\mrm{had}_h \over m_h} = a(g_0)\,,
\ees
for any value of the bare coupling $g_0$. At this point the 
theory is parametrized in terms of physical 
observables namely $\nf+1$ hadron masses. 
A new observable, e.g. the B-meson decay constant,
$\Fb$, relevant for flavour physics, is now a 
function~\footnote{
With $\phi_\mrm{B}=a\Fb$, the 
equivalent dimension-less form is 
$$
 {\phi_\mrm{B} \over \mu_\mrm{p}} = 
 R_\mrm{B}(\mu_\mrm{p},{\mu_\pi\over\mu_\mrm{p}},\ldots)\,,
$$
and $\Fb=m_\mrm{p}R_\mrm{B}(0,{\mu_\pi\over\mu_\mrm{p}},\ldots)$
is the continuum prediction.
}
\bes
   \Fb = \Fb(a,\mprot,\mpi,\ldots)\,. \label{e:param}
\ees
With the typical approximation of neglecting the 
u--d quarkmass difference as well as the influence
of the top-quark on low--energy QCD, a useful 
set to be chosen in \eq{e:param} is 
\bes
        \mprot=938.3\,\MeV  \,,\;
        \mpi=139.6\,\MeV  \,,\;
        \mk=493.7 \,\MeV \,,\;
        \md=1896\,\MeV  \,,\;
        \mb=5279 \,\MeV\,. \nonumber
\ees
We observe that this experimental input is very precise.
Using it as the starting point, the running renormalized quark masses and
coupling can be computed (at high momentum scale)
as well as other quantities; in particular 
matrix elements needed for flavour physics, 
such as $\Fb$.

Of course the continuum limit $a\to0$ has to be taken to arrive at the
physical quantity, $\Fb(0,\mprot,\mpi,\ldots)$. 
The only assumption  is that this limit exists.
(In practise, it is taken by a numerical extrapolation of a few data points
at different $a$.) There is a large amount of evidence,
perturbative and non-perturbative and also from the investigation
of critical phenomena, that given a local action
(that means a discretization with exponentially localized
interactions) the continuum limit does exist. The perturbative 
renormalization group self-consistently predicts  
that the continuum limit is at $g_0=0$: $a(0)=0$.

\subsection{Some results}

In recent years more and more results have appeared with rather 
small quoted error bars. We refer the reader to 
reviews by M. Della Morte and A. J\"uttner at
``Lattice 2007'' \cite{lat07:michele,lat07:andreas} and, 
focusing more on recent results,
the one of E. Gamiz at ``Lattice 2008'' \cite{lat08:gamiz}. 
In \tab{t:gamiz} we collected 
a few quantities, interesting for flavour physics, 
for which errors around 1\,\% are quoted \cite{lat08:gamiz}.

\begin{table}[htb!] 
\begin{center}
\begin{tabular}{ llll }
\hline
$\mcharm^{\msbar}(3\,\GeV) $ &=& $ 0.986(10)\,\GeV $ & HPQCD \cite{mc:hpqcd}\\
$\mbeauty^{\msbar}(\mbeauty) $ &=& $ 4.20(4)\,\GeV $ & HPQCD \cite{lat08:gamiz}\\
$\xi={\Fbs\sqrt{\mbs}\over\Fb\sqrt{\mB}} $ &=& $ 1.211(38)(24) $ & FNAL/MILC \cite{fbs:fnal}\\
$\Fbs $ &=& $ 243(11)\,\MeV $ & FNAL/MILC \cite{fbs:fnal}\\
$\Fds $ &=& $ 241(3)\,\MeV $ & HPQCD \cite{fds:hpqcd}\\
\hline
\end{tabular}
\caption{Results discussed in the review of E. Gamiz \protect\cite{lat08:gamiz}.
\label{t:gamiz}} 
\end{center}
\end{table}

Taken at face value, the last line in the table is in 
disagreement with what the CLEO experiment finds; see the talk of 
Philip Rubin at this conference. 
Should we conclude
in favour of new physics \cite{dsdecays:Kronfeld}, or is this
precocious? As a preparation for an answer to this 
question, we take a rough look at the machinery used
in practise to obtain the results.

\subsection{The machinery}\label{s:machinery}

At this point in time, the numbers with the smallest quoted error bars come
from so-called rooted staggered fermion simulations 
(all those in \tab{t:gamiz} do). 
There are serious concerns about the validity of this formulation,
see for example 
\cite{rooting:con}. Since a local action
equivalent to rooted staggered fermions is not known, 
the arguments
justifying the rooting trick represent an involved discussion  
based on additional assumptions.~\cite{lat06:steve,rooting:pro}

Also NRQCD is used for the b-quark in the analysis
which yields \tab{t:gamiz}. As an effective
field theory it contains power divergences;
this means terms that diverge as $1/a^n$.
In lattice NRQCD they are subtracted
perturbatively. Uncancelled divergences of the 
form  
\bes \label{e:powdiv}
  {g_0^{2(k+1)} \over a\;  m_b}
        \sim {1 \over a\, [\log(a)]^k \mbeauty} \;\toas{a\to0}\; \infty\,
\ees
are then left behind when one works until $k$'th order in perturbation
theory.
The continuum limit therefore does not exist and the estimate
of the precision of extracted results is delicate.
Other sophistications commonly employed include
smearing, many parameter fits 
(in the case of staggered chiral perturbation theory)
and Baysian fits.

All in all the large amount of tricks employed in
the computations 
compromises their directness and ``first principles status''.
One could also say systematic errors might be underestimated.
It is often emphasized that 
the methods are tested  by a comparison to experimental numbers. 
In our opinion, this is appropriate for gaining a rough idea
on the validity range of model calculations, but not
the way to establish a controlled theory computation. 
Before concluding that new physics is present, 
it appears very desirable to perform independent computations with an
independent technology. In particular we insist on
\bi
\vspace*{-2mm}\item a manifestly local discretization and
\vspace*{-2mm}\item non-perturbative subtraction of power law divergences.
\ei
\vspace*{-2mm}Such computations are in progress. Before turning to
them, let us understand why they were not carried out
already a number of years ago. The basic reason is that lattice 
QCD is a considerable 
challenge.

\section{The challenge}\label{s:chall}
This becomes apparent by considering just 
the typical hadronic input \eq{e:param}.  There is a large range of
scales between $\mpi\approx 140\,\MeV$ over
$\md=2\,\GeV$ to $\mb=5\,\GeV$. In addition, the
 ultraviolet cutoff of $\Lambda_{\rm UV} = a^{-1}$
of the discretized theory has to be large compared to all 
physical energy scales if the discretised theory is to
be an approximation to a continuum. 
Also the linear extent of space time has to
be restricted to a finite value $L$   
in a numerical treatment: 
there is an infrared cutoff $L^{-1}$.
Together the following constraints have to
be satisfied.
\bes
     \Lambda_{\rm IR} \;=\;  L^{-1} & \ll\;m_\pi\,,\;\ldots\;,\md\,,\mb\;\ll &
     a^{-1} \;=\; \Lambda_{\rm UV}  \nonumber \\
\mbox{effects} \qquad\qquad\qquad        \rmO(\rme^{-L\mpi})& & \rmO((a\, m^\mrm{had})^2)  \nonumber \\  \nonumber
        \downarrow && \downarrow \\ \nonumber
        L \gtrsim 4/\mpi \approx 6\,\fm && a \lesssim 1/(2\md)\approx
        0.05\,\fm \qquad\qquad   \\[-1ex]
\mbox{yielding} \qquad\qquad\qquad\qquad\quad       & L/a \gtrsim 120\,. & \nonumber
\ees
After the first line, we have discarded the scale of mesons
or baryons with b-quarks. Thus these are not included in the 
already rather intimidating estimate of $L/a \gtrsim 120 $. 
Looking at these numbers,
it appears unavoidable to separate the b-quark mass scale
from the others before simulating the theory. We shall return to this
in \sect{s:beauty}.

\section{Perspectives}\label{s:persp}
Despite the considerable challenge that we are
facing, there are good perspectives for 
meeting it soon. First, without being able to 
enter into the merited in-depth discussion, I mention
that the condition $L\gtrsim 6\,\fm$ may be relaxed by choosing 
unphysically large pion masses combined with a subsequent 
extrapolation. A factor of 1.5 to 2 may be gained this way.
Such extrapolations are theoretically guided 
by chiral perturbation theory~\cite{chpt:gl1},
also including lattice spacing effects~\cite{chpt:ss,chpt:asq}.

\begin{figure}[t!]
\begin{center}
\psfig{figure=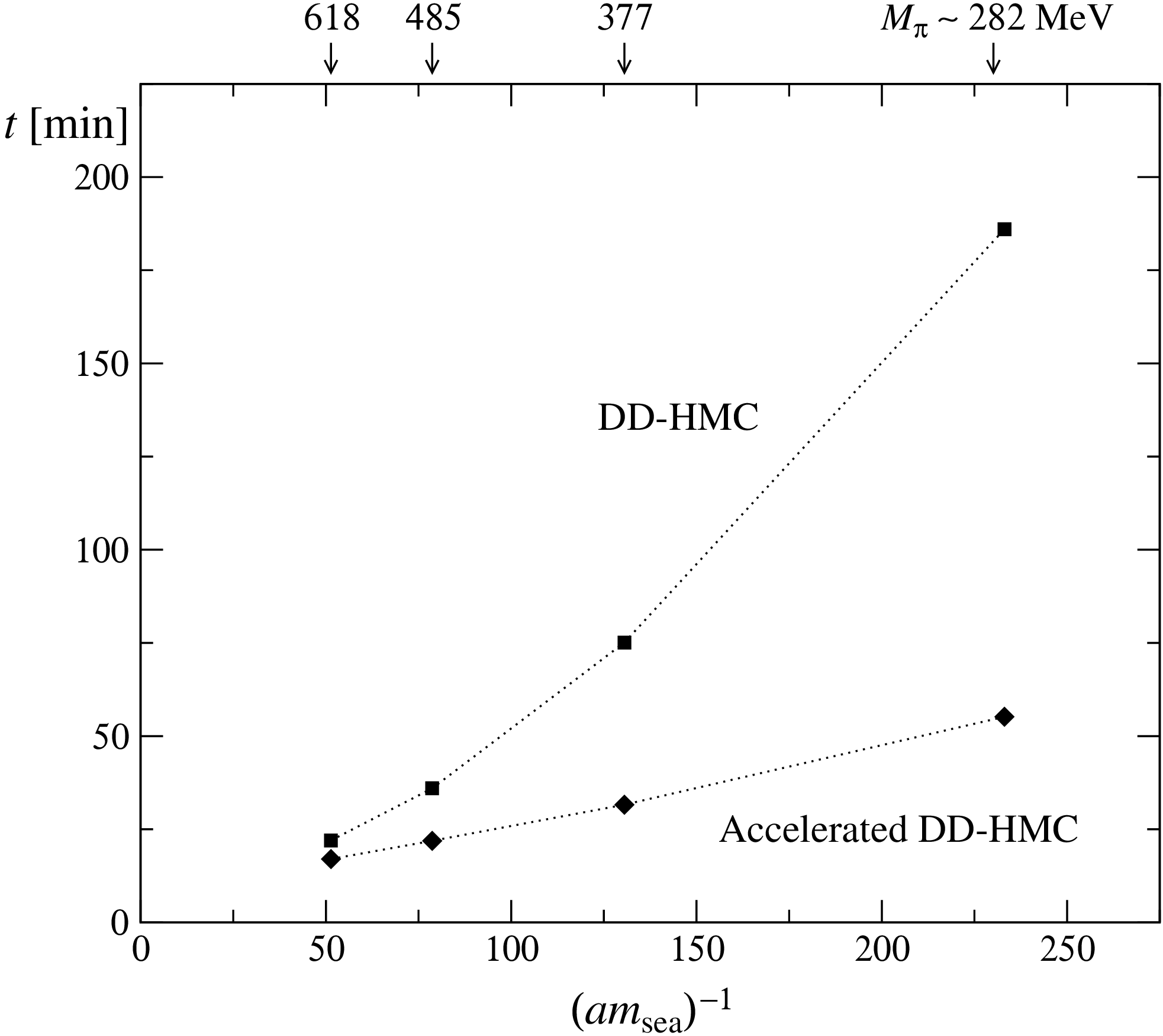,height=6.5cm}
\caption{Performance of DD-HMC as a function of the inverse
  quark mass \protect\cite{algo:L3}. 
``Accelerated DD-HMC'' refers (amongst others)
to the use of deflation. Note that the cpu-time $t$ for original 
HMC algorithm scales with $(am_\mrm{sea})^{-n}\,,\; n\geq 3$.\hfill
\label{f:perf}}
\end{center}
\end{figure}

\subsection{Algorithms}\label{s:algo}
Second, it is not obvious what is the CPU-effort
for a QCD simulation. Indeed, since the 
development of the first exact algorithm called
Hybrid Monte Carlo (HMC)
\cite{hmc}, it has been improved 
considerably~\footnote{Even though we list only buzzwords
here, it is impossible to do justice to all relevant
developments and the interested reader is advised to
consult reviews \cite{nara:tony}.} by multiple time-scale integration 
\cite{Sexton:1992nu,algo:urbach}, the Hasenbusch 
trick of mass-preconditioning 
\cite{algo:GHMC,algo:GHMC3}, the use of the domain decomposition
method in QCD 
\cite{algo:L1,algo:L1a,algo:L2} and a deflation method \cite{algo:L3}
combined with chronological inverters \cite{algo:chrono}. In parallel,
parameters of the algorithms may be tuned \cite{algo:trajlength}
and it has been understood that instabilities due to the 
breaking of chiral symmetry become irrelevant in sufficiently large volumes
\cite{algo:stability}. Also algorithms for the contributions of
single flavours 
(not mass-degenerate within a doublet) have been developed
\cite{algo:PHMC,algo:RHMC}. It is further suggested that 
fluctuations at small quark masses can be tamed by clever reweighting
techniques \cite{algo:reweight:stefan,lat08:filippo}. 

Amongst all these developments, we emphasize
the recent taming of the critical slowing down with the quark
mass. It was achieved by M. L\"uscher through a combination of
deflation with others of the tricks listed above. The impressive
weak dependence of the execution time on the quark mass 
is illustrated in \fig{f:perf} taken from~\cite{algo:L3}.

\subsection{Machines}\label{s:machines} 
It is well known that the speed of computers has  been 
increasing continuously -- at an exponential rate. Yet, it is 
illustrative to look at an (incomplete and personal)  
list of the machines available
over the years. Tflop numbers have to be taken with care,
but the overall message on the fast growth of available
computer power is clear. Because of {\em recent} investments into
high performance computing, the growth has been stronger than Moore's law.

\begin{table}[htb!]
\begin{center}
\begin{tabular}{lcrl}
\hline
 year & machine   & speed/Tflops & share for a typical \\[-0.5ex]
 & & & LQCD  collaboration \\[1ex]
1984 & Cyber205  & 0.0001       & /100  \\
1994 & APE100    & 0.0500       & /4 \\
2001 & APE1000   & 0.5000       & /4 \\
2005 & apeNEXT   & 2.0000       & /3 \\ 
2009 & PC-Cluster ``Wilson''& 15.0000     & /1 \\
2009 & BG/P      & 1000.0000 & /20(?) \\
\hline
\end{tabular}
\caption{An incomplete list (in particular between 
1984 and 1994 there were a number of other machines) of computers on which we 
have done QCD computations. \hfill \label{t:mach}}
\end{center}
\end{table}

\begin{figure}[t!]
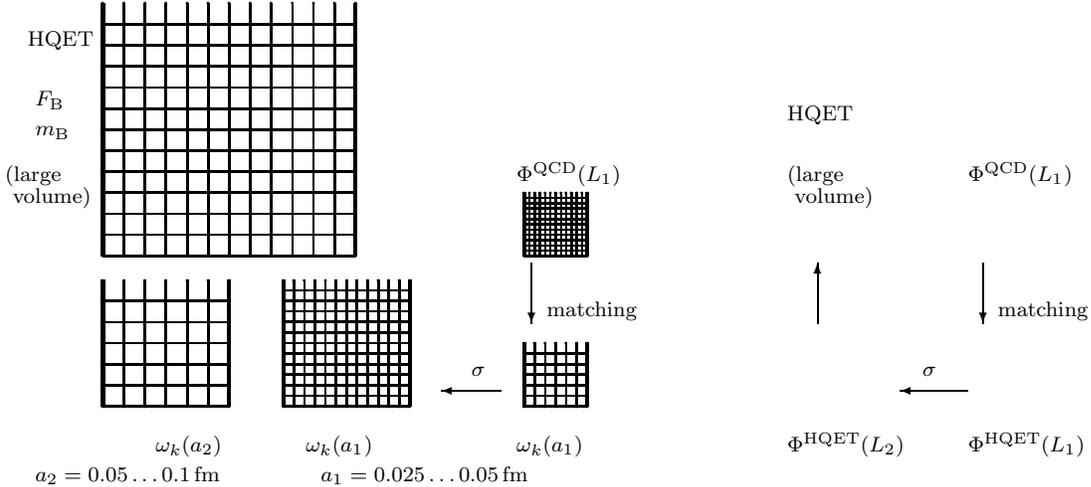

\input plots/hqetstrat12.tex
\caption{HQET strategy. The parameters in the HQET Lagrangian and
fields are denoted by $\omega_k$. In the matching step they are directly 
determined from requiring $\Phihqet_i(a,L_1,\mbeauty)=\Phiqcd_i(a=0,L_1,\mbeauty)$ 
(other masses are supressed). 
In a second step 
they are used to compute $\Phihqet_i(a=0,L_2=2L_1,\mbeauty)$ by an 
extrapolation in $a_1$. Returning to finite $a$,  
allows to extract $\omega_k(a_2)$  at larger lattice spacings $a_2$. These
are sufficient for infinite volume computations. The r.h.s. shows the same
procedure understood as a relation between renormalized quantities. \hfill
\label{f:hqet}}
\end{figure}

\subsection{Strategies for the b-quark}\label{s:beauty}
B-meson decays and mixing are an essential part of flavour physics. 
We have seen above that their direct, ``relativistic'',  treatment  
is impossible, since the demand of $L/a \gtrsim 120$ would be increased
by yet another factor of 2-3. Conversely we may, however, also
use the large mass of the b-quark to our advantage, working in an
expansion in $\Lambda_\mrm{QCD}/\mbeauty$. There are kinematical
limitations to this expansion: momenta of other particles also have
to be small compared to the b-quark mass. This means for example
that the larger part of phase space in a decay $B\to\pi l \nu$ 
is {\em not} accessible to HQET, the effective field theory which 
implements the expansion. In the analysis of the experimental
data an apropriate cut has to be implemented.

The most significant difficulty in a non-perturbative use of HQET
is, however, of a different origin. We already mentioned power law
divergences in \sect{s:machinery}. These appear in the effective theory
and have to be subtracted non-perturbatively, which means the parameters
of the effective theory have to be determined non-perturbatively, thus
avoiding terms of the form \eq{e:powdiv}. In 
contrast to QCD, new parameters appear
at each order in the HQET expansion. Still, their determination follows
the logic as described around \eq{e:masses}. There is only one twist
to the general principle: instead of 
using a relatively large number of {\em experimental} inputs, which 
results in a loss of predictive power~\footnote{For example it is hard 
to imagine how one could avoid to have the decay constant among the 
input instead of among the predictions.}, one may {\em compute} quantities 
in lattice QCD with a relativistic b-quark and use those as input.
This step is denoted by non-perturbative matching.
It was realized already a while ago~\cite{lat01:rainer,hqet:pap1} that such a 
strategy can be 
carried out by considering quantities defined in a finite volume
of linear dimension around $L=0.2\ldots 0.5\,\fm = \rmO(1/\Lambda_\mrm{QCD})$
in this initial step.
Some technical obstacles
had to be removed~\cite{stat:letter,stat:actpaper,hqet:pap3,gevp:pap}, 
but now the path appears paved for a precise 
implementation of the idea. The strategy is depicted in \fig{f:hqet}.

An impression of the achievable precision may be gained from
the pioneering computation \cite{hqet:pap3}, where 
for the first time $\minv$ corrections were taken into account. 
We warn the reader that this was a quenched computation, where
sea-quark effects are ignored. It should be used to judge
on the achievable precision, not for input into phenomenology. 
The computation focussed on the
relation between the (spin-averaged) B-meson mass and the renormalization
group invariant b-quark mass, $\Mbeauty$. \Tab{t:Mb} displays the results
for $\Mbeauty$ at the lowest order in $\minv$. Units are the potential
scale~\cite{pot:r0} $r_0\approx0.5\,\fm$. The second column of the table
lists the results in the static approximation, where terms of order
$\Lambda$ in $\Mbeauty$ are included, while corrections of
order $\Lambda^2/\mbeauty^2$ relative to the leading term 
are omitted. Since the computation is carried out with different
matching conditions, specified by an angle $\theta_0$ appearing in the 
boundary conditions for the quarks, their spread of about 3\% can
be taken as an indication for the size of the corrections. At the next order
in $\minv$ (columns 3--5), even more different  matching conditions are 
explored (angles $\theta_1,\theta_2$). The spread of results is now
reduced to a level of below 1 part in 200, much smaller than the total
error of the result~\footnote{Such a statement can be made since errors
for different entries in the table are signficantly correlated.}. 
We conclude that precision computations are possible in HQET. 

\input table_Mb.tex

As a preparation for a discussion of a second method for implementing 
the scale
separation, let us turn to the right part of \fig{f:hqet}. 
There the procedure is interpreted as a relation 
\bes
  \Phiqcd(a_0,L_1,\mbeauty) = \Phihqet(a_1,L_1,\mbeauty)
  \;\to \Phihqet(a_1,L_2,\mbeauty) = \Phihqet(a_2,L_2,\mbeauty) 
  \qquad \nonumber \\ 
  \qquad \qquad \qquad \qquad \;\to  \mbox{large volume physics} 
  \nonumber
\ees
The fact that only observables appear in this chain, with the parameters
in the Lagrangian of the theory eliminated,  
makes clear that the continuum limits $a_i\to 0$ can be taken in each step.
In practise, $a_0$ down to $0.01\,\fm$ is used while 
 $a_2$ ranges between $0.05\,\fm$ and $0.1\,\fm$.

The second method starts from the same idea and the same (or similar)
quantities in $L=L_1$. But then it remains in relativistic QCD, computing 
\bes
  \sigma_i(L_1) = \lim_{m_\mrm{h}\to\mbeauty} \lim_{a_1 \to 0} 
  {\Phiqcd_i(a_1,L_2,m_\mrm{h}) \over \Phiqcd_i(a_1,L_1,m_\mrm{h})}\,,
\ees
for each desired observable $\Phiqcd_i$. 
Similarly one connects to the large volume in the 
last step \cite{romeII:fb,romeII:mb}.
The advantage is that there are no truncation errors 
of order $\minv^2$. On the downside, lattice spacings 
$a_1,a_2$ have to be considerably smaller and/or the masses 
$m_\mrm{h}$ in the numerical extrapolation $m_\mrm{h}\to\mbeauty$ 
have to be significantly smaller than the physical one. However, the
combination $L_i m_\mrm{h}$ is always large compared to one
and therefore a smooth extrapolation of the functions
$\sigma_i$ in $1/ (L_i m_\mrm{h})$ is expected and found numerically.
A most promising approach is to combine the advantages of the two 
methods \cite{ssm:comb1}. In fact, let us compare
\bes
\mbar^{\msbar}_{\beauty}(\mbar_\beauty)&=& 4.347(48) \GeV  \nonumber
   \qquad\mbox{HQET~\cite{hqet:pap3}} \\[-1ex] &&  
   \qquad\qquad\qquad\qquad\qquad\qquad\qquad\qquad \mbox{quenched!} \nonumber
   \\[-1ex]
\mbar^{\msbar}_{\beauty}(\mbar_\beauty)&=& 4.421(67) \GeV \nonumber
   \qquad\mbox{combination~\cite{ssm:comb1}} 
\ees       
The comparison is meaningful because the experimental inputs 
(which matter  in the quenched approximation) have been chosen the same.
Let us finally note that the dominating uncertainty here is the 
one in the renormalization factor for the quark mass 
in the relativistic theory~\cite{alpha:nf2,mbar:nf2}; this can be improved.

\begin{figure}[htb!]
\psfig{figure=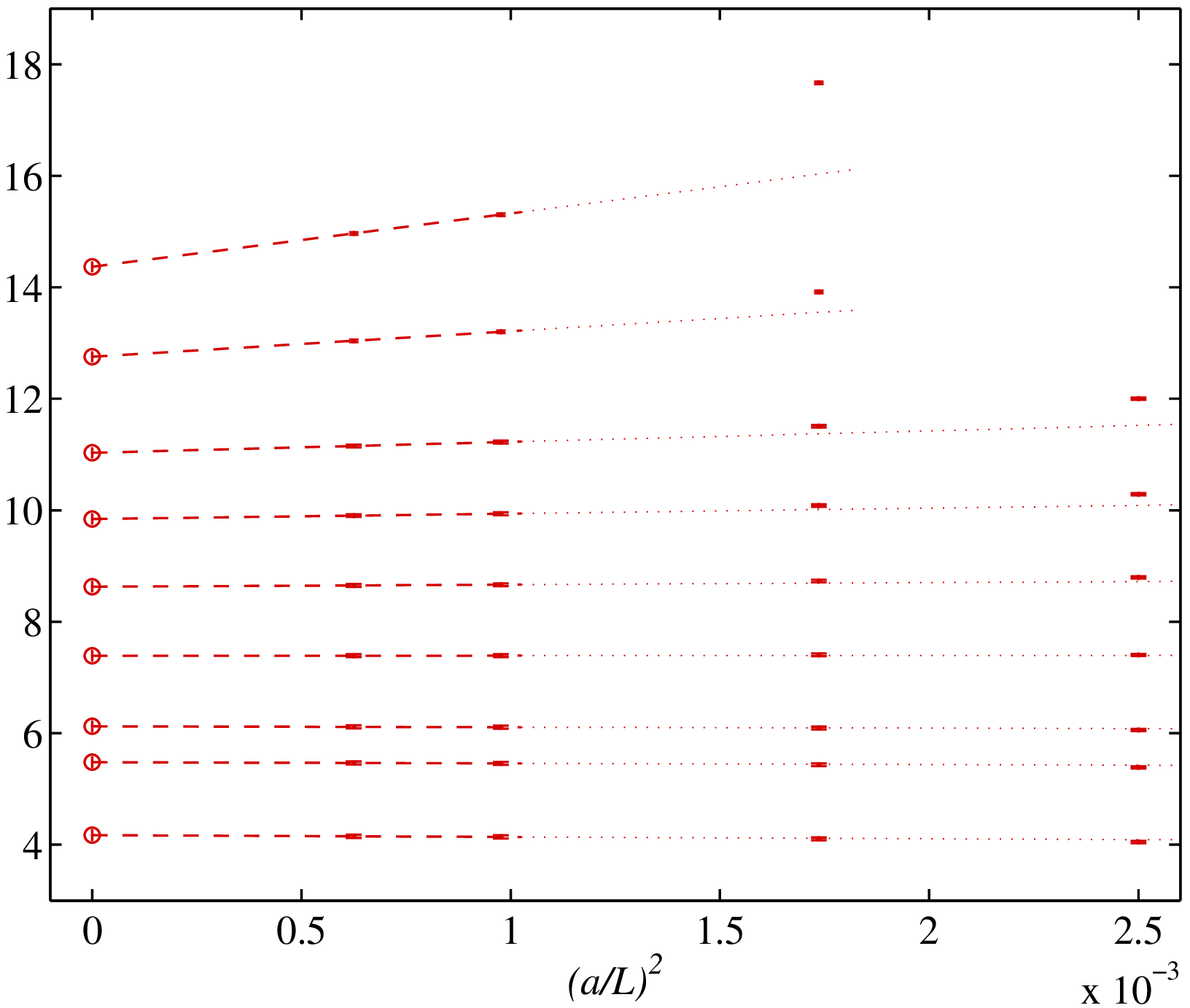,height=6.5cm}\hfill
\psfig{figure=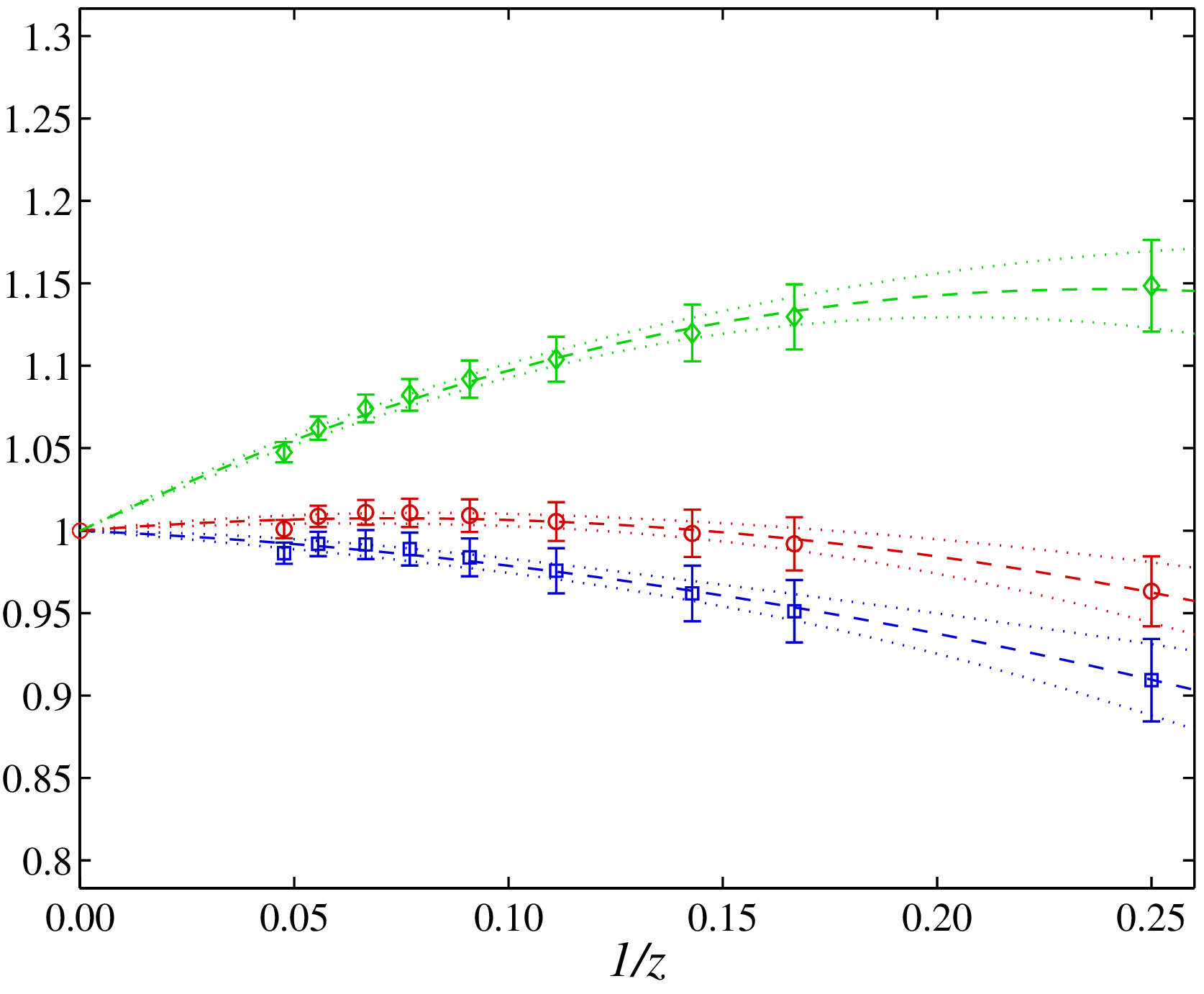,height=6.5cm}
\caption{Example of a continuum limit $\lim_{a_0\to0} \Phiqcd(a_0,L_1,\mbeauty)$
 for a whole range of $\mbeauty$ (left) and the resulting mass dependence
 on the right \protect\cite{lat08:patrick}. The variable $z$ is the 
dimensionless RGI quark mass $z=\Mbeauty L$.  
 Curves are a low order HQET-representation. For details
 see the reference.  \hfill
\label{f:hqettest}}
\end{figure}

We emphasize that dynamical fermions are absolutely needed 
for a comparison to phenomenology.
Presently, these strategies are being applied to $\nf=2$ QCD 
(just up and down sea). An intermediate result~\cite{lat08:patrick}
shown in \fig{f:hqettest} demonstrates the precision of the 
most difficult continuum extrapolations.

\subsection{The CLS strategy}

While the numerical effort for the physically small lattices
(see \fig{f:hqet}) is quite modest for today's computers, 
the theoretical setup is to be chosen with care. In particular
the \SF is used~\cite{SF:LNWW,SF:stefan1} where Dirichlet boundary
conditions in time play a prominent r\^ole. This again calls for
a simple, very local action. 

Such an action, the $\Oa$-improved Wilson 
action~\cite{impr:SW,impr:pap1,impr:csw_nf2}
is then also needed in the large volume part. 
The strategy of the CLS effort~\cite{CLS} is to use it 
and through its simplicity be able to profit from
developments in algorithms \cite{algo:L3}. This helps to diminish 
systematic and statistical errors by 
working at small lattice spacings and on physically large 
volumes. Furthermore perturbative errors and in 
particular divergences (such as \eq{e:powdiv}) are removed 
by non-perturbative calculations in the \SF.

The present goal are simulations with $\nf=2$ dynamical quarks,
lattice spacings 
$a=0.08,0.06,0.04\,\fm$, sizes $L=2\,\fm$ 
to $4\,\fm$ and pion masses down to $\mpi=200\,\MeV$. 
Concerning heavy flavours
both the B-physics programme 
emphasized above and charm physics~\cite{lat08:georg} 
are being investigated. 
We note that  a considerable  problem remaining
will have to be
solved along the way: the slow evolution of topological 
modes~\cite{algo:topo:luigi} in standard MC algorithms. 

\section{Conclusions}

It is now widely recognized that lattice QCD is 
an important tool for computing phenomenologically 
relevant quantities. 
I have often heard complaints that progress is slow, 
but one has to recognize  that the rich dynamics and the many relevant
scales in QCD represent quite a challenge (\sect{s:chall}). 
Despite this 
we appear to approach the era of controlled precision 
predictions in time for an accurate analysis of present and 
future flavour physics data and models.
In particular present worries about the use of the rooting trick and
uncancelled power divergences can become history.

\section*{Acknowledgments}
I would like to thank my colleagues in CLS and ALPHA
for a fruitful collaboration, and in particular U. Wolff 
and S. Schaefer for comments on the manuscript.
We acknowledge support
by the Deutsche Forschungsgemeinschaft
in the SFB/TR~09
and by the European Community
through EU Contract No.~MRTN-CT-2006-035482, ``FLAVIAnet''.
Our simulations are performed on BlueGene and APE Machines of
the John von Neumann Institute for Computing
at FZ J\"ulich and DESY, Zeuthen,
on PC-clusters at the Universities of Rome La Sapienza and 
Valencia-IFIC and at CERN, as well as on the IBM MareNostrum
at the Barcelona Supercomputing Center. 
We thankfully acknowledge the computer resources and technical support provided
by all these institutions.

\section*{References}
\bibliographystyle{h-elsevier}   
\bibliography{refs}           

\end{document}

%% file: plots/hqetstrat12.tex
\newcommand{\bla}{}
\newcommand{\cbla}{}
\newcommand{\blu}{}
\newcommand{\red}{}
\newcommand{\cred}{}
\newcommand{\cmag}{}
\newcommand{\slat}[1] 
{
\unitlength #1
\linethickness{0.2mm}
\bla\multiput(0,0)(0,2.0){6}{\line( 1, 0){12}}
\bla\multiput(0,0)(2.0,0){6}{\line( 0, 1){12}}
\linethickness{0.4mm}
\multiput(0,2)(2.0,0){7}{\bla\circle*{0.2}}
\multiput(0,4)(2.0,0){7}{\bla\circle*{0.2}}
\multiput(0,6)(2.0,0){7}{\bla\circle*{0.2}}
\multiput(0,8)(2.0,0){7}{\bla\circle*{0.2}}
\multiput(0,10)(2.0,0){7}{\bla\circle*{0.2}}
\multiput(0,12)(2.0,0){7}{\red\circle*{0.2}}
\multiput(0,0)(2.0,0){7}{\red\circle*{0.2}}
\put(0,0){\blu\line(0,1){12}}
\put(12,0){\blu\line(0,1){12}}
}
\newcommand{\blat}[1] 
{
\unitlength #1
\linethickness{0.2mm}
\bla\multiput(0,0)(0,2.0){12}{\line( 1, 0){24}}
\bla\multiput(0,0)(2.0,0){12}{\line( 0, 1){24}}
\linethickness{0.4mm}
\multiput(0,2)(2.0,0){13}{\bla\circle*{0.2}}
\multiput(0,4)(2.0,0){13}{\bla\circle*{0.2}}
\multiput(0,6)(2.0,0){13}{\bla\circle*{0.2}}
\multiput(0,8)(2.0,0){13}{\bla\circle*{0.2}}
\multiput(0,10)(2.0,0){13}{\bla\circle*{0.2}}
\multiput(0,12)(2.0,0){13}{\bla\circle*{0.2}}
\multiput(0,14)(2.0,0){13}{\bla\circle*{0.2}}
\multiput(0,16)(2.0,0){13}{\bla\circle*{0.2}}
\multiput(0,18)(2.0,0){13}{\bla\circle*{0.2}}
\multiput(0,20)(2.0,0){13}{\bla\circle*{0.2}}
\multiput(0,22)(2.0,0){13}{\bla\circle*{0.2}}
\multiput(0,24)(2.0,0){13}{\red\circle*{0.2}}
\multiput(0,0)(2.0,0){13}{\red\circle*{0.2}}
\put(0,0){\blu\line(0,1){24}}
\put(24,0){\blu\line(0,1){24}}
}

\unitlength 0.2cm
\begin{picture}(35,30)(-5,-3.5)

\scriptsize 
\put(28.5,11){\blat{0.035cm}}
\put(28.5,1){\slat{0.07cm}}
\put(12.5,1){\blat{0.07cm}}
\put(0.5,1){\slat{0.14cm}}
\put(0.5,11){\blat{0.14cm}}
\red\put(-4.5,25.0){HQET}
\red\put(-4.0,19.0){$\mb$}
\red\put(-4.0,21.0){$\fb$}
\red\put(-6.0,16.0){(large}
\red\put(-6.0,14.5){ volume)}
\red\put(15.0,-4.0){$a_1=0.025\ldots0.05\,\fm$}
\red\put(-4.0,-4.0){$a_2=0.05\ldots0.1\,\fm$}
\red\put(4.0,-2.0){$\omega_k(a_2)$}
\red\put(14.0,-2.0){$\omega_k(a_1)$}
\red\put(28.0,-2.0){$\omega_k(a_1)$}
\cbla\put(29,10.5){\vector(0,-1){4.0}}
\cbla\put(30.0,7.0){matching}
\cmag\put(28.0,16.0){$\Phiqcd(L_1)$}
\cred\put(27,2){\vector(-1,0){4.0}}
\put(25,3){$\sigma$}

\red\put(46,20.0){HQET}
\red\put(46.0,16.0){(large}
\red\put(46.0,14.5){ volume)}
\red\put(46.0,-2.0){$\Phihqet(L_2)$}
\red\put(58.0,-2.0){$\Phihqet(L_1)$}
\cbla\put(48,6.5){\vector(0,1){4.0}}
\cbla\put(59,10.5){\vector(0,-1){4.0}}
\cbla\put(60.0,7.0){matching}
\cmag\put(58.0,16.0){$\Phiqcd(L_1)$}
\cred\put(58,2){\vector(-1,0){4.5}}
\put(55,3){$\sigma$}

\normalsize
\end{picture}


%% file: table_Mb.tex
\begin{table}
\begin{center}
\begin{tabular}{cccccc }
\hline
$\theta_0$ &  $r_0\,\Mb^{(0)}$ && \multicolumn{3}{c}{$r_0\,\Mb=r_0\,(\Mb^{(0)} + \Mb^{(1a)} + \Mb^{(1b)})$} \\
\hline
& &&
$\theta_1=0$   &  $\theta_1=1/2$ & $\theta_1=1$ \\
& &&
$\theta_2=1/2$ &  $\theta_2=1$   & $\theta_2=0$ \\
\hline
 && \multicolumn{4}{c}{Main strategy} \\
0   & 17.25(20) && 17.12(22)  & 17.12(22)  & 17.12(22) \\
\hline
    &           & \multicolumn{4}{c}{Alternative strategy} \\
0   & 17.05(25) && 17.25(28)  & 17.23(27)  & 17.24(27) \\
1/2 & 17.01(22) && 17.23(28)  & 17.21(27)  & 17.22(28) \\
1   & 16.78(28) && 17.17(32)  & 17.14(30)  & 17.15(30) \\
\hline\scriptsize
 & 3 \%& \multicolumn{4}{c}{$0.6\% \ll$ total error} \\
 & $=\rmO(\Lambda^2/\mbeauty^2)$ & 
\multicolumn{4}{c}{$=\rmO(\Lambda^3/\mbeauty^3)$} \\
\hline
\end{tabular}
\caption{ \label{t:Mb} Renormalization group invariant b-quark mass. $\Lambda\equiv\Lambda_\mrm{QCD}$.}
\end{center}
\end{table}
